# Sensitivity and robustness of Lagrangian coherent structures in coastal water systems

Anusmriti Ghosh[1], K. Suara[1], Scott W. McCue[2], Richard J. Brown[1]

[1]Environmental Fluid Mechanics Group, Queensland University of Technology (QUT), QLD 4000, Australia

[2]School of Mathematical Sciences, Queensland University of Technology (QUT), QLD 4000, Australia

## Abstract

In coastal water systems, horizontal chaotic dispersion plays a significant role in the distribution and fate of pollutants. Lagrangian Coherent Structures (LCSs) provide useful tools to approach the problem of the transport of pollutants and have only recently been applied to coastal waters. While the fundamentals of the LCS approach using idealised analytical flow fields are well established in the literature, there are limited studies on their practical implementations in coastal waters where effects of boundaries and bathymetry frequently become significant. Due to their complex bathymetry and boundaries, unstructured grid systems are commonly used in modelling of coastal waters. For convenient derivation of LCS diagnostics, structured grids are commonly used. Here we examine the effect of mesh resolution, interpolation schemes and additive random noise on the LCS diagnostics in relation to coastal waters. Two kinematic model flows, the double gyre and the meandering jet, as well as validated outputs of a hydrodynamic model of Moreton Bay, Australia, on unstructured grids are used. The results show that LCSs are quite robust to the errors from interpolation schemes used in the data conversion from unstructured to structured grid. Attributed to the divergence in the underlying flow field, the results show that random errors in the order of 1-10 % cause a break down in the continuity of ridges of maximum finite time Lyapunov exponents and closed orbit elliptic LCSs. The result has significant implications on the suitability of applying LCS formulations based on a deterministic flow field to diffusive coastal waters.

## Highlights

- The work examines the sensitivities of applying Lagrangian coherent structures (LCSs) diagnostics for coastal waters with complex boundaries.
- LCSs are robust to the errors from interpolation schemes used for unstructured to structured grid velocity data conversion.
- Additive random errors in the order of 1-10 % cause a break down in the continuity of ridges of maximum finite time Lyapunov exponents and closed orbit elliptic LCSs.

## 1. Introduction

Coastal waters form a transitional dynamic zone between the sea and land, and have a significant environmental, social, and economic benefit. Management of these coastal waters and estuaries is more challenging with increasing pollution pressures [1]. The major sources of contaminants or pollutants in coastal water systems are urbanisation, industrialisation, mining, agriculture, coastal catchment development, population growth [2,3] and climate change [3]. Due to the complex flow fields of these coastal water systems, the pollutant transport and accumulation is complex [4,5]. Identifying the source and prediction of the transport pathway of pollutant particles in coastal water system is an important challenge.

Hydrodynamic modelling has a noteworthy impact for the transport of sediments and pollutants in these coastal water systems [6]. Generally, unstructured rather than structured mesh are used in hydrodynamic models [7,8]. Coastal water systems have complex boundaries, thus unstructured or flexible meshes are comparatively more accurate than a structured mesh [9].

LCSs are defined by the locally strongest attracting and repelling material lines over a given time interval [10]. The idea of LCSs provides a new way of understanding transport in complex fluid flows and a way to organise material transport by identifying key material lines [11]. LCSs have received significant attention in the last decades due to their usefulness in different areas, such as transport and mixing of turbulent flow properties [12] and pollutant transport in the atmosphere and ocean [13-16]. They have been extensively used for studies in the ocean and large water bodies to understand a range of problems [16]. For example: Bettencourt et al. [17] studied 3D oceanic LCSs in the upwelling region of Benguela using the finite-size Lyapunov exponent (FSLE) detection method; d'Ovidio et al. [18] used LCSs in the Mediterranean Sea to understand the transport and mixing structures; Lekien et al. [15] observed LCSs on the Florida coast to reduce industrial contamination effects; Huhn et al. [19] used LCSs in the South Indian Ocean to study the Madagascar plankton bloom and found that advection is important; and Prants [20] discussed LCSs in relation to mixing and transport in the ocean. Most publications apply LCSs to understand large water bodies or ocean problems but very few studies can be found on shallow water estuarine systems [21].

LCSs can be diagnosed by several approaches which are discussed in detail by Hadjighasem et al. [22]. The majority of these approaches require the calculation of flow maps from particle trajectories. This formulation can be achieved using both structured and unstructured grids [23,24]. However, for ease of computation in many engineering packages, e.g. MATLAB, structured grid formulations are commonly used [25,26]. As hydrodynamic model outputs of coastal waters are mostly available with unstructured grids, there is a need for data conversion from unstructured to structured grids to perform the relevant LCS analysis. This data conversion process requires selection of an appropriate interpolation scheme and mesh resolution combinations that minimise error effects on the computed LCSs.

The fundamental approach of LCSs using idealised analytical flow fields is well established, but the practical implementation of the identification approaches in coastal waters where effects of boundaries and bathymetry frequently become significant are limited. The effect of data conversion (unstructured to structured grid for LCS calculation), sensitivity of interpolation schemes and robustness of LCSs identification to velocity errors in the underlying flow field is less well understood. Haller [23] discussed spiky velocity errors with a certain time interval and concluded that in oscillating perturbations, LCSs can be robust even with significant amplitudes. Shadden et al. [27] observed errors for the hourly radar-derived surface velocity data and showed that LCSs are robust to noise. Harrison, Glatzmaier [28] examined the average velocity error effects using a unique random noise at the height of weekly sea surface fields and found that LCSs are relatively less sensitive than the Okubo–Weiss (OW) parameter. The specific knowledge of random error is not clear for each and every point of the periodic velocity field flow. Although LCSs are quite robust to the errors in approximate discrete velocity field data [27,23,28] the level of noise when breakdown of LCS identification occurs is still unclear.

Grid transformation and random noise are major sources of uncertainty in hydrodynamic modelling that could inherently affect the application of LCS to coastal waters. Therefore, the aim of the present study is to (i) examine the effect of errors associated with data conversion on LCS diagnostics, (ii) examine the sensitivity of LCS diagnostics to mesh resolution and standard interpolation approaches used in data conversion, and (iii) examine the robustness of LCSs to different magnitudes of additive random noise. In this study two idealised analytical flow fields are first used to examine the effects of grid data conversion and random noise because these flow fields are devoid of uncertainty except those imposed on them. We then extend the analyses to a real domain to ascertain the validity of our findings in real coastal water system. Two different types of analytical kinematic flows (double gyre and meandering jet) are used to examine the effect of mesh resolution, interpolation scheme, and noise on the LCSs field. Furthermore, we present a case study using a hydrodynamic model output to further examine the effect of mesh resolution, interpolation scheme and analysis of LCSs robustness. The case study location is Moreton Bay, a semi enclosed tidal embayment, in southeast Queensland, Australia.

## 2. Material and Methods
### 2.1 Description of velocity field data

In the present work, we utilised two analytical kinematic models and pre-validated hydrodynamic model outputs. The analytical kinematic models are used because they are deterministic in space and time, require no validation and are idealised physical flow patterns such as vortices observable at different timescales in coastal waters. Furthermore, hydrodynamic model output of a real system is used as a case study to show validity to the real flow field. The flow field from the hydrodynamic model of Moreton Bay is validated against extensive field observations using fixed and moving instruments. The model description and validation are fully detailed in [6] and applied LCS analysis in [29] thus, only brief description is provided in this paper.

#### 2.1.1 Analytical Kinematic models

Open and closed kinematic models are considered in this investigation, namely the meandering jet and double gyre, respectively. As the meandering jet is of the open flow field type, it is an idealised representation of water bodies such as estuarine systems. The closed double gyre flow field on the other hand is an idealised representation of water bodies such as lakes, where basin-scale vortices are initiated by wind forcing. For the analyses presented in this paper, the analytical kinematic equations will be described using an unstructured grid containing 5,000 points generated randomly within the domain. A uniform random distribution (discrete) is used through the MATLAB software to create the unstructured grid velocity field. The selection was such that grid points cannot overlap within the domain.

##### 2.1.1.1 Double gyre

The double gyre analytical kinematic model is a time-dependent two-gyre system which contains two counter rotating vortices [30]. The flow system of the double gyre is used to investigate many shallow and closed flow fields [31,32]. This unsteady kinematic model can be described by a stream function [33] via

$$\psi(x, y, t) = A\sin(\pi f(x, t)) \sin(\pi y), \tag{1}$$

where

$$f(x, t) = a(t)x^2 + b(t)x, \tag{2}$$

$$a(t) = \epsilon \sin(\omega t), \tag{3}$$

$$\text{and} \quad b(t) = 1 - 2\varepsilon \sin(\omega t). \tag{4}$$

In this study, we are interested in the time-dependent gyres, thus $\epsilon \neq 0$. From Equation (1), the velocity field can be extracted by taking the partial derivatives of the stream function:

$$u = -\frac{\partial \psi}{\partial y} = -\pi A \sin(\pi f(x)) \cos(\pi y)$$

$$v = \frac{\partial \psi}{\partial x} = \pi A \cos(\pi f(x)) \sin(\pi y) \frac{df}{dx}.$$

(5)

Following Shadden et al. [33], the parameter values that have been used in this study are $A = 0.1$, $\omega = 2\pi/10$, $\epsilon = 0.25$, where the flow timespan is $t \in [0, 10]$ and the domain is $x \in [0, 2]$, $y \in [0, 1]$ [34]. The timespan is discretised using equally spaced [0,10] time steps. The timespan is selected for the oscillation period of the counter-rotating vortices.

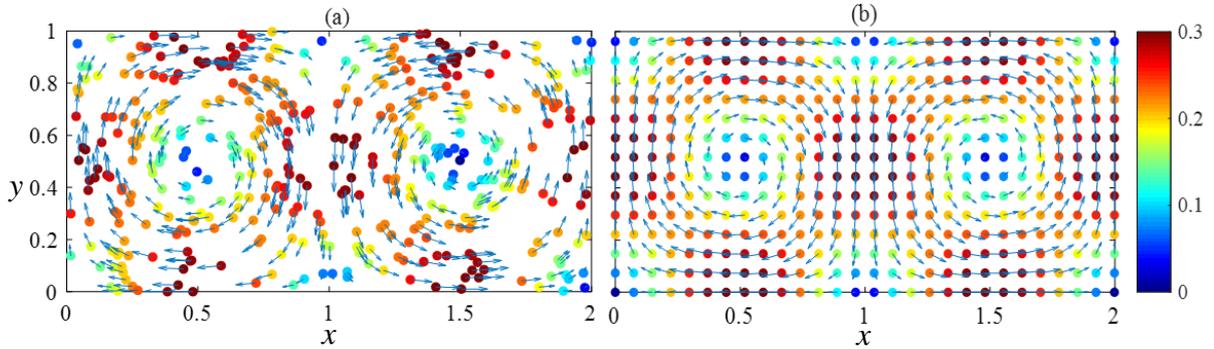

**Fig. 1:** Double gyre quiver plot coloured by velocity magnitude at $t = 10$ for (a) unstructured and (b) structured grid data showing velocity vectors (approximately 400 grid points).

Fig. 1 shows the velocity vector (Equation (5)) using these parameters on the unstructured and structured grid containing approximately 400 points (for clear visualisation) coloured by velocity magnitude at $t = 10$.

### 2.1.1.2 Meandering jet

The meandering jet kinematic model system contains two dynamically distinct recirculation regions (eddies). These two eddies are separated by the flow trajectory of a meandering jet [28]. This model divides the dynamics of the fluid into distinct sections and provides a framework which helps to understand particle transport [12]. The flow system of the meandering jet can be used to investigate many coastal and open ocean flow fields [28,35]. The kinematic model of the meandering jet can be described by a stream function [28,36]

$$\psi = -by + A\sin(x - \omega t)\sin(\pi y), \qquad (6)$$

where *x* and *y* are along the jet and across the jet horizontal coordinates, respectively [28]. From the stream function, a two - dimensional, incompressible and time dependent velocity field can be extracted by taking the partial derivatives of the stream function [28,37]:

$$u = -\frac{\partial \psi}{\partial y} = b - A\sin(x - \omega t)\cos(y)$$
$$v = \frac{\partial \psi}{\partial x} = A\cos(x - \omega t)\sin(y)$$
(7)

The parameter values that are used in this study are $A = 0.5$, $\omega = 0.04$, $b = 0.25$, where the flow timespan is $t \in [0, 10]$ and domain is $x \in [0, 6]$, $y \in [0, 3]$. These parameter values are chosen following Harrison, Glatzmaier [28]. Through these selected parameters the meandering jet eddies are moving to the right as shown in Fig. 2. The timespan is selected to reveal the complete cycle of meandering jet flow pattern. The meandering jet flow coherent structures generation and evolution with time are discussed detail in Flierl et al. [38].

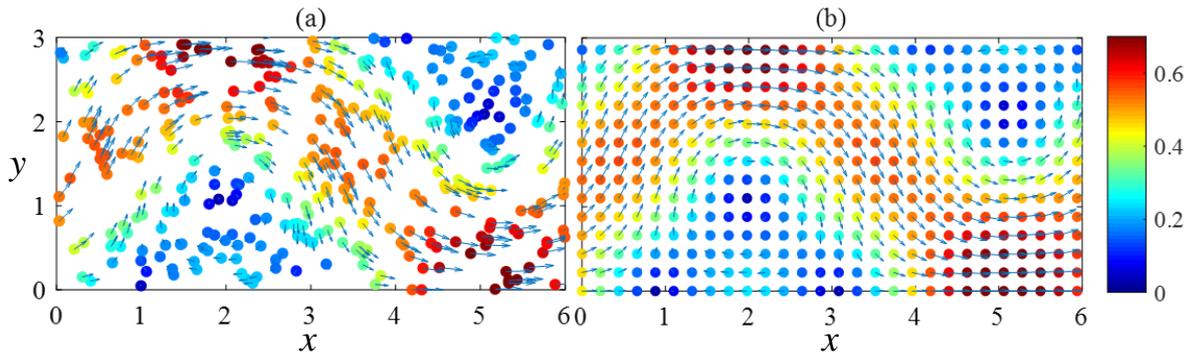

**Fig. 2:** Meandering jet coloured by velocity magnitude at $t = 10$ for (a) unstructured and (b) structured grid data showing velocity vector (containing approximately 400 points).

Fig. 2 shows the velocity vectors on the unstructured and structured grid containing approximately 400 points (for clear visualisation) coloured by velocity magnitude at t = 10. The fluid recirculates in circular orbits around the centres of the eddies near the fixed point where the velocity vanishes for both unstructured and structured grid points. On the other hand, fluid close the centre of the jet is transported downstream.

### 2.1.2 Hydrodynamic model dataset for Moreton Bay

The surface velocity data of Moreton Bay was obtained from a validated hydrodynamic model [6]. Moreton Bay is a semi-enclosed subtropical embayment in southeast Queensland. The system lies between 27º and 28º south latitude, spans approximately 110 km north to south, and has its major opening to the ocean of approximately 15 km on the northern side (Fig. 3).

The 3D hydrodynamic model was developed using MIKE3D, which has been extensively used in studying in estuaries and coastal waters systems. This hydrodynamic model set up period was between July 23 and August 6, 2013 (Fig. 3) and the time duration was chosen to overlap the period for model validation with the available field observation drifters data [6]. The horizontal domain was a network of flexible unstructured triangular grids, consisting of 13,918 elements (Fig. 3b).

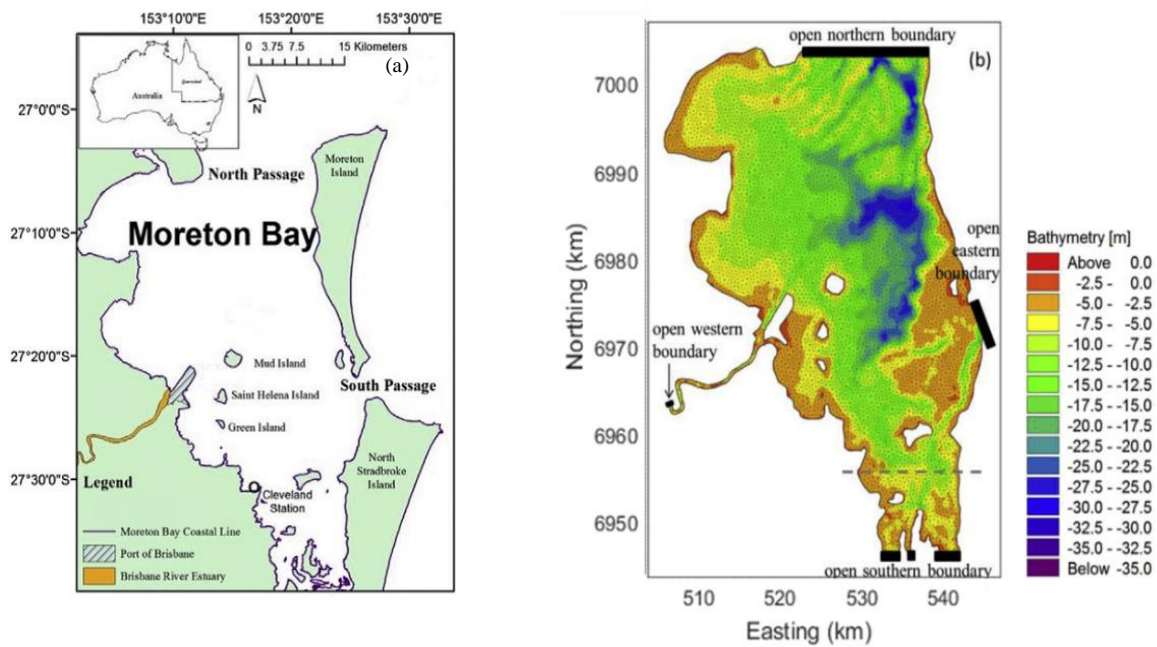

**Fig. 3:** Moreton Bay (a) map of shoreline and (b) bathymetry (Yu et al., 2016).

The coordinates (Northing and Easting) of Moreton Bay are in Universal Transverse Mercator (UTM). Near the river mouth and coastal region, a fine (<100 m) grid resolution was used. In the far-field areas, a relatively coarser grid resolution was used (i.e., 100m to 500 m) (Fig. 3b). In the vertical direction, 10 variable sigma co-ordinates layers were used [6]. Hourly river discharge data observations from the Department of Environment and Resource Management, Queensland, Australia, were used at the west boundary for the boundary condition. Ten-minute interval tidal elevations data provided by Maritime Safety Queensland, Australia, served as boundary conditions for the open northern, eastern and southern boundaries. Wind data sourced from the Australian Bureau of Meteorology at a chosen site (153.24 ºE, 27.26 ºS) at 1-minute intervals and was used as the input to the model domain [39,6]. The normalised RMSE between the observed (drifter) and trajectories calculated from the hydrodynamic model were 1.26% and 7.45% in the northing and easting directions, respectively [6]. This indicates that the model produced accurate flow field representing the dynamics of Moreton Bay. The detailed description of the spatiotemporal variation of the flow field in Moreton Bay is discussed in [6]. The analyses here were carried out using surface velocity output (top layer) with a 15-minute time interval on the unstructured grid.

## 2.2 Method of Analysis
### 2.2.1 LCS diagnostic using FTLE:

The key parameters for detection of LCSs are the flow map and the resulting Cauchy-Green strain tensor [40,30]. Upon the estimate of these parameters, different detection approaches can be applied. The calculation of the flow map and LCS depend on the nature of the flow field and seeding particles. Finite-time Lyapunov exponents (FTLE) are used in this analysis as a proxy to diagnose the hyperbolic LCSs. Although there are some limitations with the use of ridges of FTLE as a proxy for LCS [40], application of this approach has been shown to be reliable for periodic dynamical systems [29]. The FTLE field is extensively used in laminar and turbulent flows as a criterion to reveal the hidden structures in fluid flow [25].

The FTLE algorithm starts with the computation of the flow map that takes an initial fluid particle from a position $x_0$ at time $t$ to its later position $t + \tau$ in time $x(t + \tau; t, x_0)$ [41]. The FTLE fields are calculated using the following equation [41,23,22]

$$FTLE_{t_0}^{t_1}(x_0) = \frac{1}{|t_1 - t_0|} \log\sqrt{\lambda_2(x_0)}. \qquad (8)$$

In Equation (8), $\lambda_2$ is the maximum eigenvalue of the Cauchy–Green deformation tensor, $t_1$ represents the final time, and $t_0$ represents the initial time. When $t_1 > t_0$, the FTLE field represents the forward / repelling LCSs and if $t_1 < t_0$, the FTLE field indicates the backward/ attracting LCSs. To reveal the repelling/stable and the attracting/unstable manifolds particles are moving forward ($t_1 > t_0$) and backward ($t_1 < t_0$) in time correspondingly. The maximum expansion rate is represented by stable or repelling manifolds lines, whereas the maximum compression rate is represented by unstable or attracting manifold lines. Herein we only focus on the computation of the forward FTLE field, the ridge of which reveals the maximum repelling material lines i.e., the stable manifold. Particles that are next to the stable manifold lines are considered maximally repelling along those lines. The computation of the FTLE in this study was performed using scripts modified from the Barrier Tool [49] and our in-house code [29].

### 2.2.2 LCSs diagnostic using the geodesic principle

The generalized method of geodesic principles is used for the calculation of elliptic LCSs. Haller and Beron-Vera investigated fluid regions which were surrounded by exceptional material loops [42] and found a typical exponential stretching loop in turbulent flow [43,44]. According to Haller and Beron-Vera, such typical loops are identified as elliptic LCSs [10,44]. An elliptic LCS is defined as a closed orbit of coherent vortices in the fluid flow and can be seen as closed lines that ideally permit no transport of material across them. Therefore, they can be considered an idealized transport barrier that separates materials being carried by the underlying flow [45,13,46,43,47].

Elliptic LCSs loops are uniformly stretched by the same $\lambda$ factor under the flow advection time from initial to final [44]. Where $\lambda$ is a constant stretching ratio and defined as the ratios of eigenvalues in the two-dimensional flow map [48]. The key ingredients to identify the coherent Lagrangian vortices are velocity data and the Cauchy-Green strain tensor. The computation for the elliptic LCS in this study was performed using scripts modified from the Barrier Tool [49].

## 2.3 Parameters
### 2.3.1 Mesh resolution

Here, analytical flow fields are used to investigate the effects of mesh resolution and interpolation schemes on LCS detection. Generally, satisfactory model accuracy depends on high-resolution data [50]. The first parameter considered is the effect of mesh resolution used for the unstructured to structured grid transportation on the LCS. Here we investigate how the mesh resolution affects the hyperbolic LCS through FTLE. To do this, the resolution of an unstructured velocity field is kept constant whereas the resolution of the structured velocity field is varied to generate different levels of the resolution ratio (R). The resolution ratio is calculated such that R = RU/RS, where RU is the number of unstructured grid points and RS is the structured grid data. For the kinematic models (double gyre and meandering jet), the unstructured data have been created using a uniform discrete distribution from the analytical model as described in Section 2.1.1.

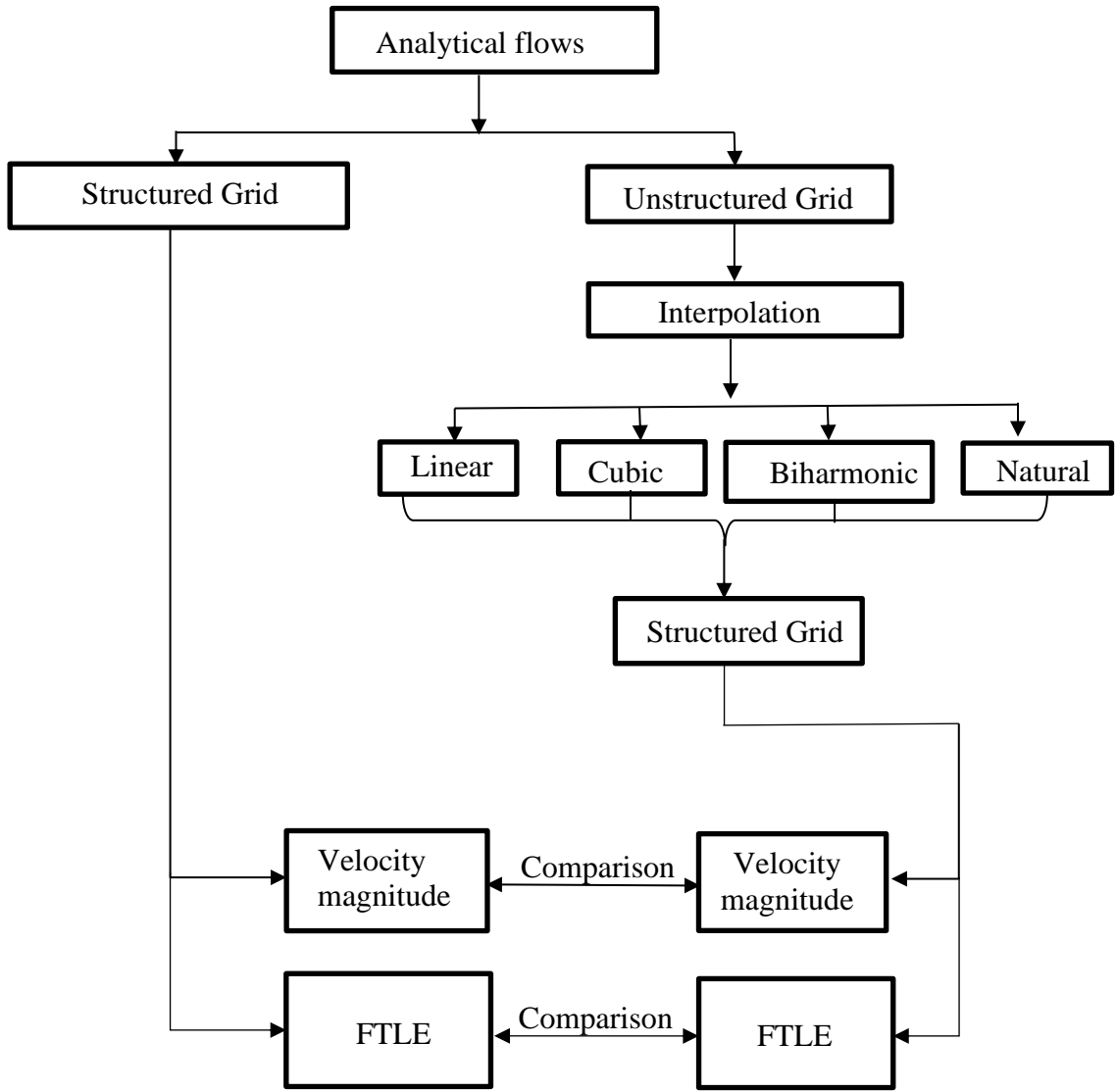

**Fig. 4:** Flow chart for the procedure used to investigate the effect of interpolation scheme and mesh resolution (R = RU/RS) on the FTLE.

Fig. 4 shows the detailed procedure implemented for different mesh resolutions to transform from unstructured to structured grid velocity and estimate the FTLE fields. In order to convert the velocity field from an unstructured to a structured grid, four interpolation schemes (linear, cubic, biharmonic and natural) were employed. The FTLE field were then calculated for these different interpolation schemes and were compared with the FTLE field computed directly from the analytical equations with equivalent RS grid resolution.

### 2.3.2 Interpolation Scheme

Different interpolation schemes are compared here to investigate their effect on hyperbolic LCS. The optimum interpolation technique is identified from the four interpolation schemes (linear, cubic, biharmonic and natural). The linear interpolation method works by interpolating a value at a query point based on the linear interpolation of values of neighboring grid points in each respective dimension [51,52]. This interpolation method depends on the specific triangulation of the data points [53]. The cubic method works by interpolating a value at a query point based on the cubic interpolation values of neighboring grid points in each respective dimension. For a smooth interpolation, based on computational time and easy implementation point of view, the cubic interpolation scheme is a good choice for data conversion [54]. The natural interpolation method is an efficient tradeoff between the linear and cubic interpolation schemes. This natural interpolation method defines the neighboring point of non-uniformly distributed data [55]. The biharmonic approach is unlike other methods, as it is not based on a triangulation [56], but instead is related to a radial basis function interpolant. This method performs the same calculation for all points regardless of location. The biharmonic interpolation scheme works best when the number of dimensions is less than three [56]. The interpolation schemes usually depend on the fit of the data characteristics, the required curve smoothness, considerations of speed, analysis of post-fit requirements, and so on. The linear and natural interpolation methods are much faster than cubic and biharmonic, however the resulting curves do not follow the grid data as well. The cubic and biharmonic interpolation methods are computationally slower, however the resulting curves follow the grid data very well. The aim here is to select the interpolation schemes that result in minimal uncertainty in the velocity magnitude and LCSs field when compared with the LCSs obtained from the direct analytical equations (Fig. 4). However, care must be taken to avoid excessive computational time. Here the uncertainty is defined as the difference between the FTLE fields which are obtained from the converted structured velocity grid data and that directly obtained from the analytical velocity field grid data where the resolution of the grid was the same.

### 2.3.3 Noise description

In the literature, it has been shown that LCSs are quite robust to model velocity errors [23,57]. However, the level of random noise required to break down LCS is still not clear. The aim here is to examine the impact of random noise on the LCS diagnostics using velocity field data with different magnitudes of noise added to two different kinematic model flows (double gyre and meandering jet) and Moreton Bay data. The FTLE metric is used as a proxy for hyperbolic LCS and the closed orbits of stretching ratio, $\lambda$, for elliptic LCSs. To define the noise level, the true velocity field is degraded by addition of random noise such that

$$U(x,y,t) = u(x,y,t) + k\varepsilon_1, \qquad (9)$$
$$V(x,y,t) = v(x,y,t) + k\varepsilon_2$$

where $k$ is the weighting factor, $\varepsilon_1$ is a normal distribution of zero mean and a standard deviation corresponding to $\sigma(u(x,y,t))$ and $\varepsilon_2$ is a normal distribution of zero mean and a standard deviation corresponding to $\sigma(v(x,y,t))$. Because of the periodic nature of the flow, the standard deviation of the underlying flow is selected to parameterise the noise. The recomputed velocity, $U(x, y, t)$ and $V(x, y, t)$ data are the degraded velocity field based on the noise. The magnitude of the noise varies with the standard deviation of the true velocity field. The weighting factor, $k$, has been used to control the relative magnitude of the noise and the true velocity such that $k$ varies from 0 - 2. Because the velocity magnitude varies for the flow types considered here, the noise weighting factor, $k = 0 – 2$ corresponds to average noise magnitude to velocity magnitude ratio of 0 – 65 % for double gyre and 0 – 16 % for meandering jet and 0 – 185 % for Moreton Bay.

## 3. Results and Description

Here we present the results and discussion of the effects of mesh resolution and interpolation schemes on the accuracy of the LCS diagnostics. We also discussed the sensitivity of FTLE to different degrees of random noise to examine the robustness of LCS.

### 3.1 Mesh Resolution and Interpolation effect on LCSs

In this section, the mesh resolutions and interpolation techniques are discussed for analytical flows as well as in the Moreton Bay case study. The resolution effect on FTLE is examined by varying the size of the structured grids relative to the averaged size of the unstructured grid systems. The four different interpolation schemes (linear, cubic, biharmonic and natural) that are commonly used in MATLAB program are examined.

#### 3.1.1 Analytical kinematic models

Here the effect of mesh resolution and interpolation of the velocity field on the FTLE fields are examined. For this purpose, we compare the results for the FTLE using the lowest and highest velocity flow field grid resolutions. The velocity fields used herein are obtained by directly computing the velocity from the analytical solutions. The kinematic models (double gyre and meandering jet) are both computed on the high resolution (201 x 101) and low resolution (13 x 7) meshes from their respective analytical equations (5) and (7). The FTLE are computed using a mesh grid that is five times finer than that of the velocity field. This refinement factor was obtained from a sensitivity analysis in which it was shown that the statistics of FTLE converges after four times the velocity grid [29]. The integration time of $\tau = 10$ is selected for the FTLE in the double gyre. This time corresponds to the period of oscillation of the counter-rotating vortices and is selected to represent the largest scale of oscillation in this flow field.

Similarly, the integration time $\tau = 10$ is used for the FTLE in the meandering jet. This time corresponds to the period of oscillation of the fluctuation of moving eddies. The results of the meandering jet represent a flow that is moving forward in time and the fluctuation of eddies are moving to the right of the sinous meandering jet. A further increase in the integration time does not reveal additional structures but rather increases the sharpness of the ridges of maximum FTLE and increases the overall computation time.

Fig. 5 compares the FTLE fields obtained for the low and high grid resolutions for both kinematic models (double gyre and meandering jet). The FTLE fields are observed to be sensitive to the resolution of the mesh grid even if it is computed directly from the analytical equation in Fig. 5. However, it can also be seen that the overall structure of the FTLE field for both grid resolutions of two different kinematic models is consistent. Fig. 5 (c) shows the FTLE field plotted as a function of locations in the $x$ direction at $y = 0.5$ for double gyre and meandering jet for two different grid resolutions (high and low) using the analytical solution. The result, however, shows that there is no significant difference for the general LCSs location and strength for grid resolutions (R = RU/RS) in the neighborhood of $x = 1$ for the double gyre and the neighborhood of $x = 2$ for meandering jet (Fig. 5c). It can be seen that the strength of the ridges of FTLE field from Fig. 5, are higher at the high grid resolution than low grid resolution. The magnitudes of the FTLE field at the ridges represent the strength of the LCSs and are different for the high and low grid resolutions (Fig. 5).

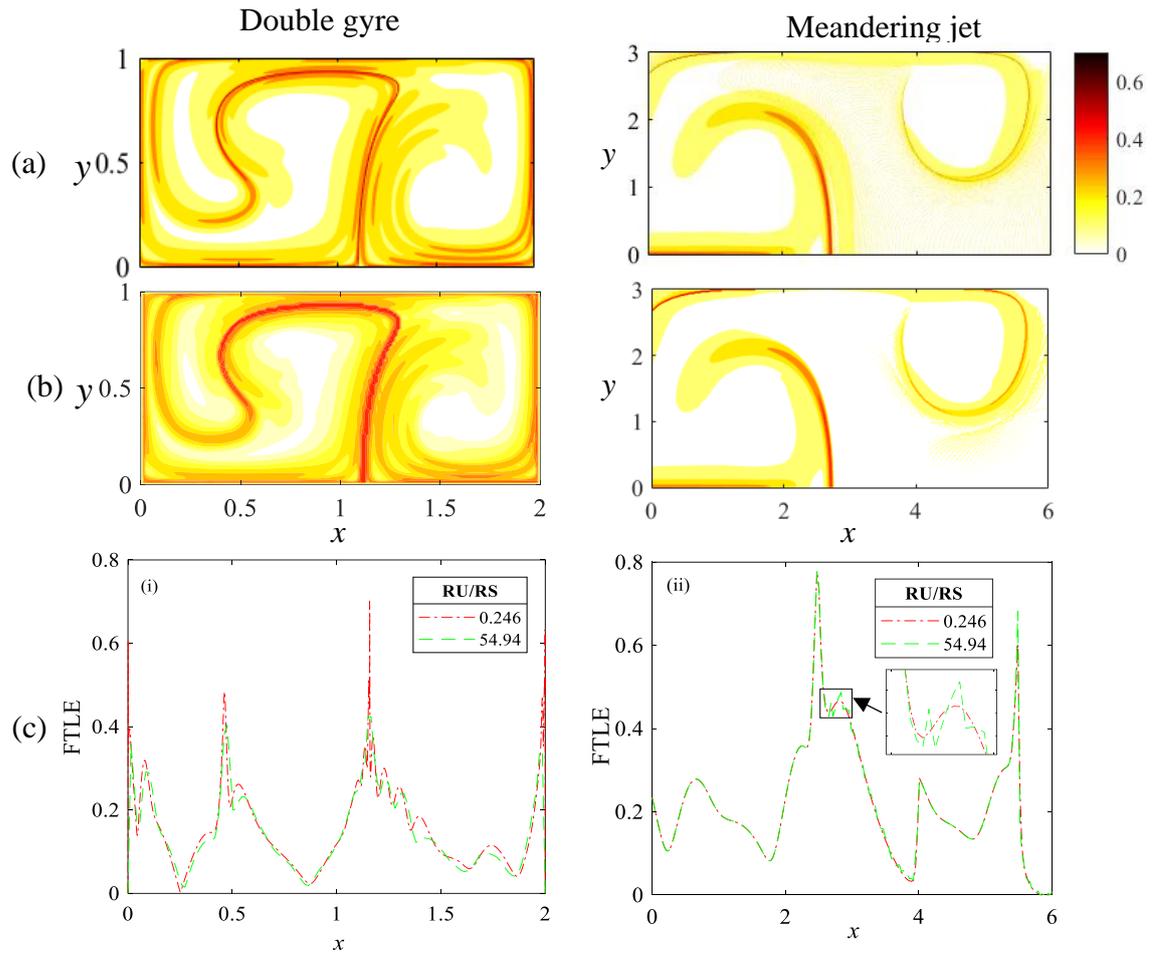

**Fig. 5:** Analytical kinematic (double gyre and meandering jet) forward FTLE fields for two different grid resolutions (RU/RS) (a) high resolution (0.246), (b) low resolution (54.94) and (c) cross section of FTLE as a function of $x$ location at $y = 0.5$ for (i) double gyre and (ii) meandering jet.

The effect of interpolation schemes in grid data conversion from unstructured to structured grid is now examined. Four different interpolation schemes are applied at two different grid resolutions (high, R = 0.246 and low, R = 54.94) to examine this effect on the results of FTLE. To examine the data conversion effect the unstructured velocity data was created using a uniform discrete random distribution technique from the both kinematic models. The number of unstructured grid points was fixed at 5,000 and the number of structured grids was varied to generate the different resolution, R. The computational time of the interpolation schemes is varied (i.e., the cubic scheme takes 2 times higher than linear scheme, while biharmonic scheme takes 5 times higher than linear scheme) as functions of the accuracy and complexity of the scheme.

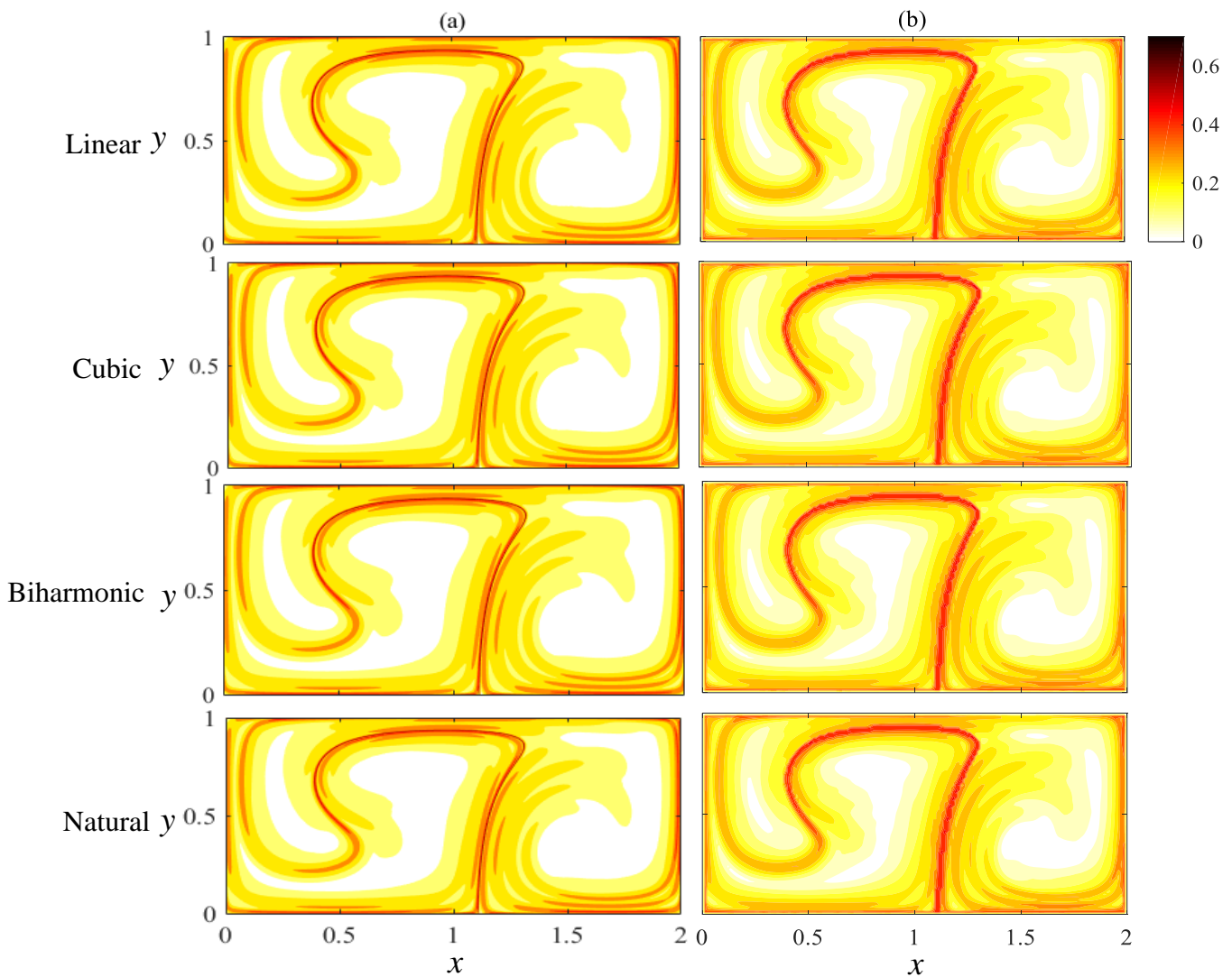

**Fig. 6:** Forward FTLE fields (colour bar) of double gyre interpolation schemes (linear, cubic, biharmonic and natural) for unstructured/structured grid resolutions (R = RU/RS): (a) R = 0.246 and (b) R = 54.94.

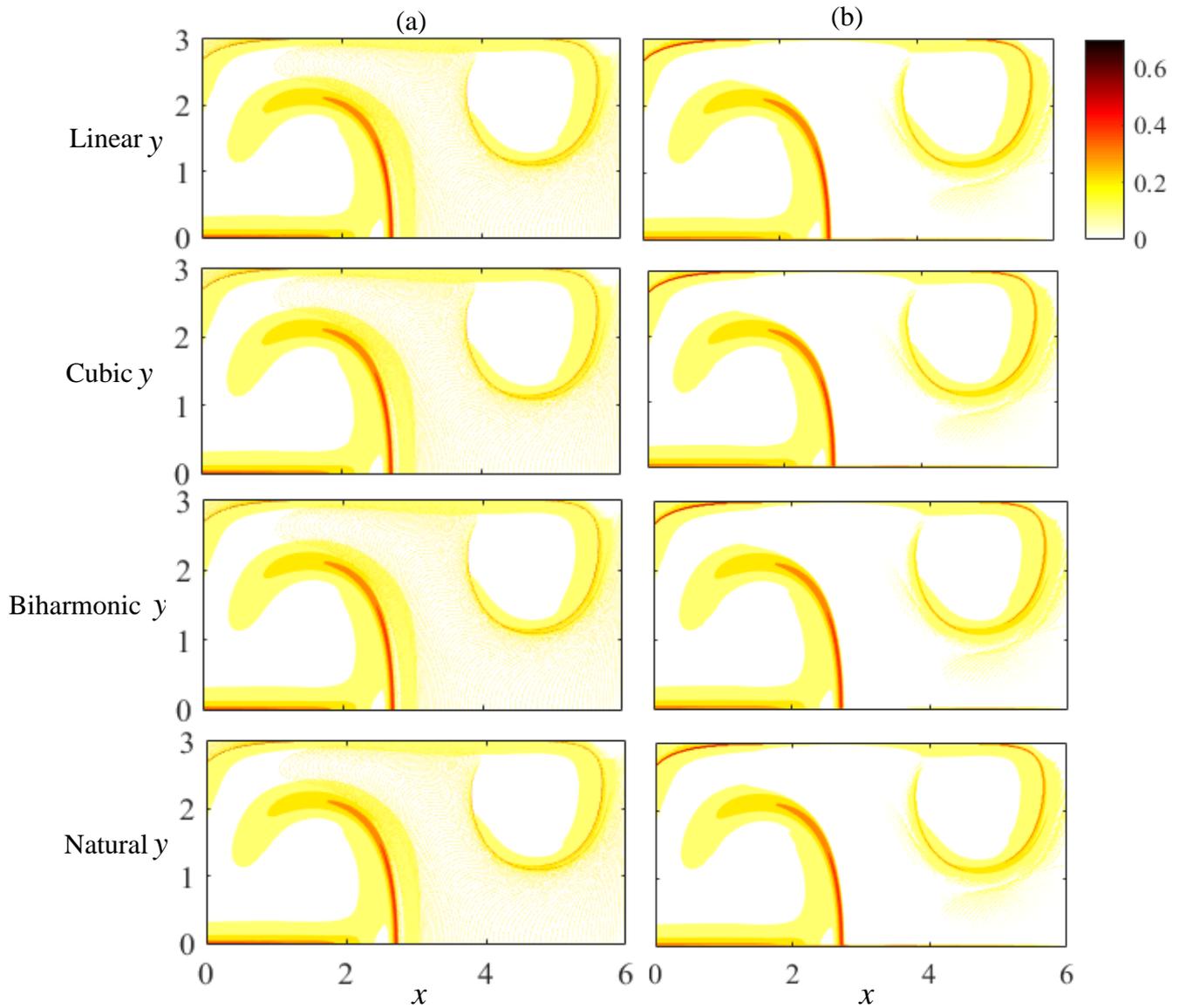

**Fig. 7:** Forward FTLE fields (colour bar) of meandering jet interpolation schemes (linear, cubic, biharmonic & natural) for unstructured/structured grid resolutions (R = RU/RS): (a) R = 0.246 and (b) R = 54.94.

The results shown in Fig. 6 and 7 are for the double gyre and meandering jet models, respectively, at low (R = 54.94) and high grid resolution (R = 0.246) with the four different interpolation schemes. The contour plots from Fig. 6 and 7 show that the LCS results for all the interpolation schemes are generally robust to random errors generated from the uncertainty in the data conversion. Small differences between Figs. 7 (a) and 7 (b) can be observed in the strength of the FTLE contour in the top left hand corner and around $x = 5$, y = 0.7, where a low level contour is persistent in Fig. 7 (b) and not in Fig. 7 (a). All Fig.s for both grid resolutions and four different interpolation schemes are qualitatively similar for both kinematic models (Fig. 6 and 7).

Fig. 8 shows the probablity density function (PDF) of the FTLE field for the (a) double gyre and (b) meandering jet, respectively. The results show that the FTLE field PDF are relatively insensitive to the interpolation schemes but varied with the different grid resolutions. The result showed that the FTLE ridges occupying the tail of the PDF (Fig. 8 (a and b)) are consistent for different interpolation scheme but not consistent with the different grid resolutions. Essentially LCSs are relatively insensitive to the interpolation schemes. These observations are consistent with results presented in Harrison, Glatzmaier [28].

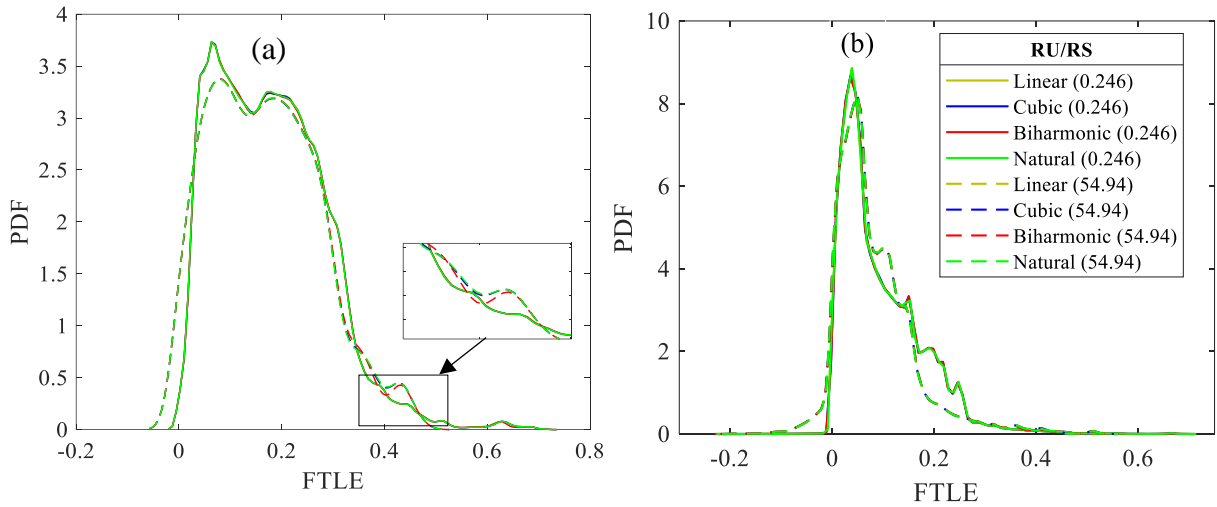

**Fig. 8:** Interpolation schemes (linear, cubic, biharmonic and natural) of highest and lowest grid resolutions for (a) double gyre and (b) meandering jet.

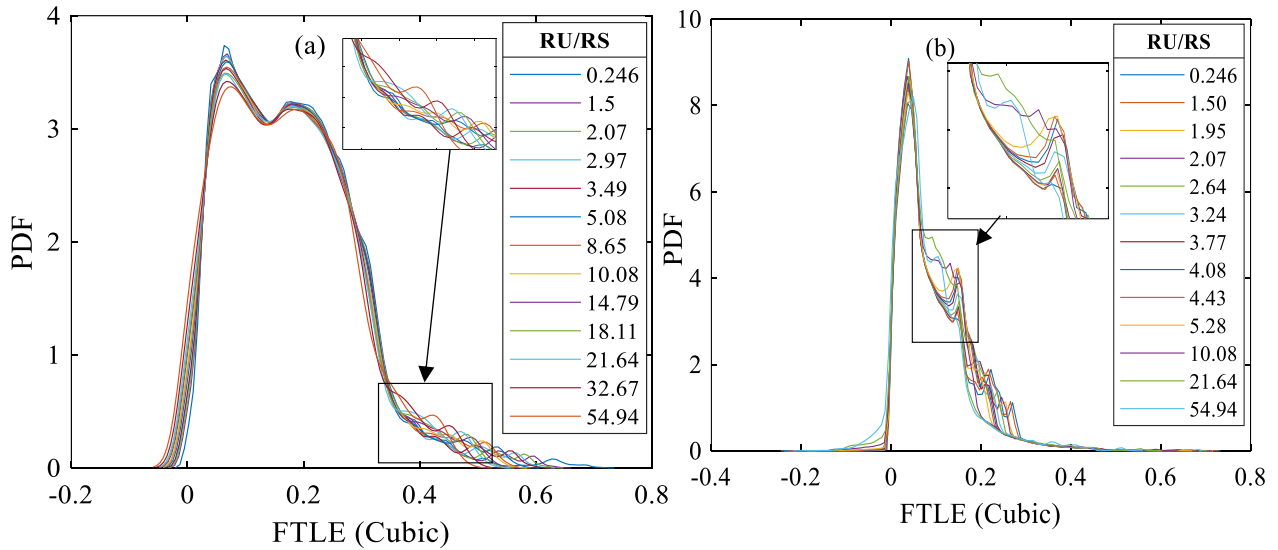

**Fig. 9:** Cubic forward FTLE field PDF for different grid resolution of (a) double gyre and (b) meandering jet.

The grid resoluion variation effects on the forwared FTLE field is shown for different grid resolutions in Fig. 9. The cubic interpolation scheme for both kinematic models was used. Fig. 9 shows that the grid resolution affects the strength of the LCS identified as the local peaks in the FTLE field with the corresponding values at the positive tail of the PDF. The strength of the LCSs clearly increases with the increasing grid resolution. The maximum and mean of the forwarded FTLE field was also examined to observe the effects of different grid resolutions in terms of different interpolation schemes in Fig. 10. The maximum FTLE field decreases with lowering the grid resolution for all the interpolation schemes. Thus, mesh resolution is an important factor for calculating LCSs. The practical implication of the results is that the flux of material passing through the LCSs can be underestimated with lower grid resolution while the location of discharge to minimize the impact of pollutants could be wrongly placed without considering an appropriate grid resolution. Therefore, further refining in the velocity field beyond the grid resolution captured in the unstructured grid does not improve the information on the LCSs field.

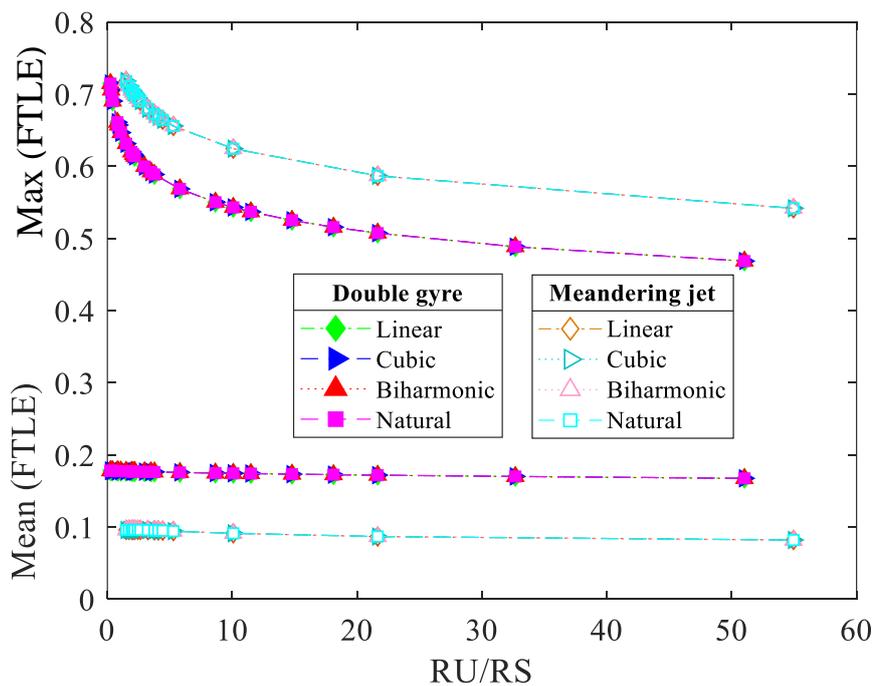

**Fig. 10:** Maximum and Mean of FTLE calculation for double gyre and meandering jet.

The results indicate that any of the interpolation schemes can be used for the unstructured – structured grid conversion. However, a close look at the boundary between the different interpolation scheme results have some variations (Fig. 6 and 7). The biharmonic interpolation scheme in the FTLE field gives the closet value to the analytical data field.

Root Mean Square Errors (RMSE) are considered between the FTLE calculated using the analytical and interpolated velocity grid data points. The RMSE is calculated as the difference between FTLE fields obtained from the analytical equation and the different interpolation velocities and is shown in Fig. 11. The uncertainty level of the FTLE field increases with an increase in the resolution (R = RU/RS), i.e., reduction in the grid size for the structured grid relative to the unstructured grid.

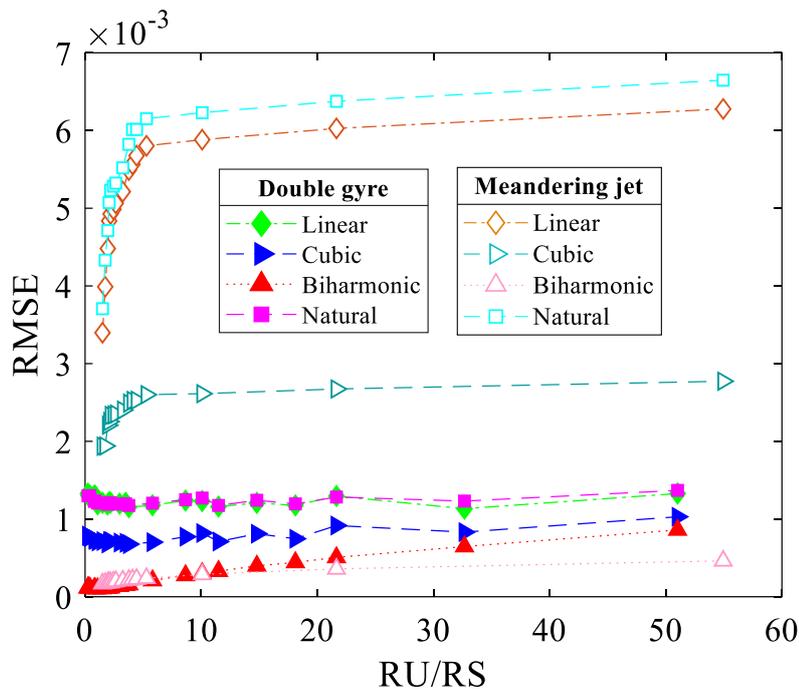

**Fig. 11**: RMSE of FTLE for different grid resolutions corresponding to four interpolation schemes (linear, cubic, biharmonic and natural) of the double gyre & meandering jet.

The RMSE result from Fig. 11 suggests that to examine LCSs in shallow water flows, the biharmonic interpolation will be the ideal choice to interpolate unstructured velocity data into a structured grid. However, due to the high computational time for the biharmonic scheme, cubic interpolation is a practical technique to interpolate the velocity data. This is because the time taken by the cubic interpolation is 10 times shorter than that of the biharmonic while the error in the cubic is at least two times lower than those obtained from Linear and natural interpolation schemes for both double gyre and meandering jets.

### 3.1.2 Case study of Moreton Bay model output

To investigate the mesh resolution and interpolation effects on field data, a tidal dominated estuary Moreton Bay was chosen as a case study. Here in, the effect of four different interpolation schemes on the forward FTLE is investigated. The output of the hydrodynamic model used in the study was obtained from a validated model of Moreton Bay discussed in detail in Yu et al. [6]. For this study, data every 15 minutes for three days velocity output on an unstructured grid was extracted for the calculation. The grid spacing for the unstructured velocity output varied between 100 - 500 m. In Fig. 12, the forward FTLE fields are computed with an integration time of 72 hours. Fig. 12 shows the FTLE field using interpolated velocity from the biharmonic interpolation technique in Moreton Bay. The contour plot of FTLE for the four different interpolation techniques revealed no qualitative difference. Thus, only the FTLE for the full domain using the biharmonic scheme is shown here in (Fig. 12). This supports the conclusion that LCSs are relatively insensitive to the interpolation scheme as was found with the analytical flows. To calculate these LCSs with interpolation schemes in Moreton Bay, the biharmonic schemes takes more than 10 times the computational time required for each of the three other interpolation schemes. To obtain the ratio of R, a small region 12 km by 20 km (shown in Fig. 12) was chosen from the full domain of Moreton Bay.

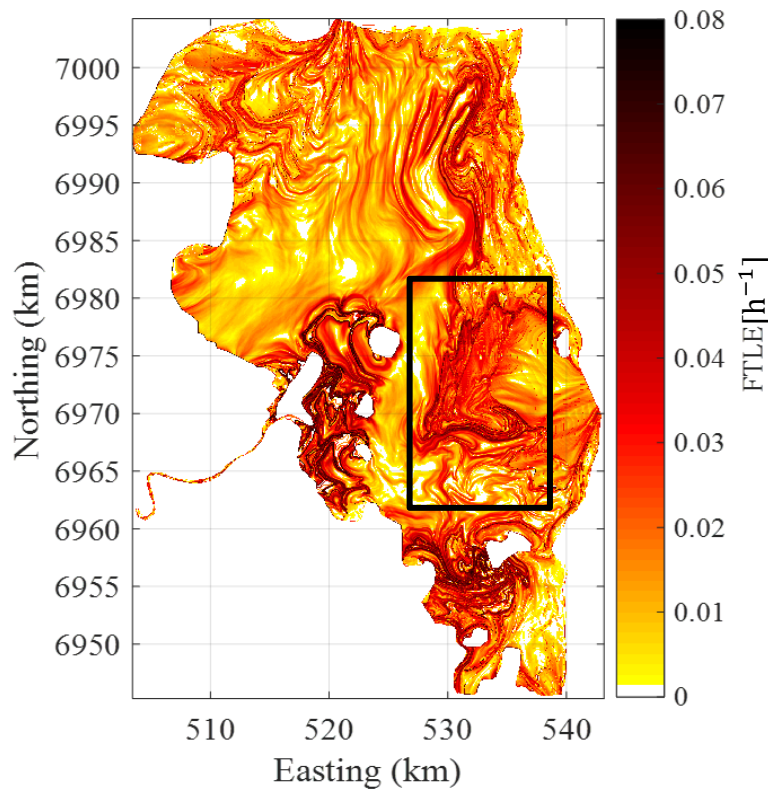

**Fig. 12:** Forward FTLE for the total domain of Moreton Bay with the analysis of biharmonic Interpolation Scheme. Rectangle shows region used to obtain resolution, R.

The number of unstructured grid points for the selected location was 1685 and it was fixed to compute different grid resolutions, R. In this field data, mesh resolution varies from R (RU/RS) = 0. 5 to 5 corresponding to (180 x 300) points and (57 x 95) points, respectively. This resolution range as selected based on the physical scale of interest constrained by the computation time and limitation of the hydrodynamic model. The converted structured velocity field of Moreton Bay is obtained using the four different interpolation techniques. The converted structured velocity field based on R, are then used in the computation of the FTLE field. For calculating the FTLE field here in for the small region, an integration time of 24 h was selected for investigating two complete semi-diurnal tidal cycles of interest.

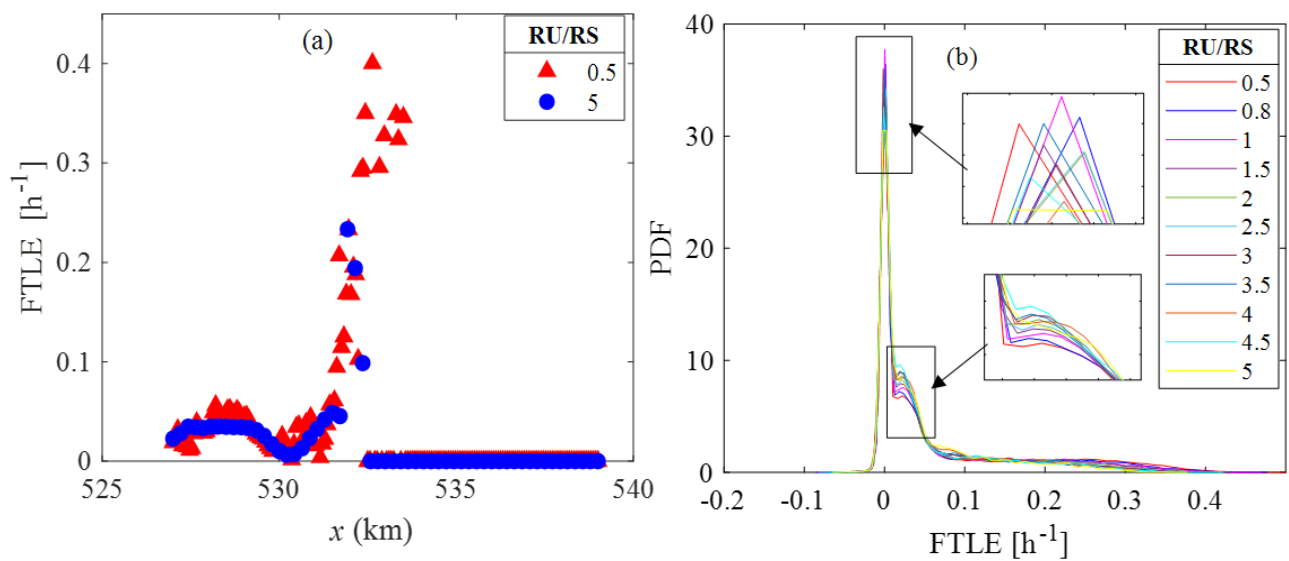

**Fig. 13:** Cubic forward FTLE calculation for different resolution of Moreton Bay (specific region) (a) $x$ location at $y = 0.5$, (b) probablity density function (PDF).

The small region in Fig. 12 was selected such that particles that are located in this region can be advected into the full domain without leaving the Moreton Bay flow field during the integration time of 24 hours. Fig. 13(a) shows the forward FTLE calculation of Moreton Bay (small region) using cubic interpolation scheme for high and low resolution at $x$ location when $y = 0.5$. Fig. 13(b) shows PDF of the forward FTLE calculation of Moreton Bay (small region) using cubic interpolation scheme for different grid resolutions. The result showed that the ridge of FTLE increases with the increase of grid resolution in Fig. 13. This is consistent with the findings for the kinematics models that proper selection of the grid resolution is required. To investigate how the maximum FTLE result changes in Moreton Bay in terms of resolution, four different interpolation schemes have been calculated in Fig. 14. This result also shows that the high grid resolution gives the highest value of FTLE which is a similar conclusion to that of the analytical flow Section. For environmental management, these results, show grid resolution can ultimately affect flux estimates through the FTLE which estimates the LCSs. If we do not account for grid resolution, then the maximum flux can be underestimated. If the flux changes, we may either overestimate or underestimate the flux through a transport barrier. On the other hand, the strength of LCSs can affect the understanding of dynamics of the system and affect the way we interpret the transport processes involved.

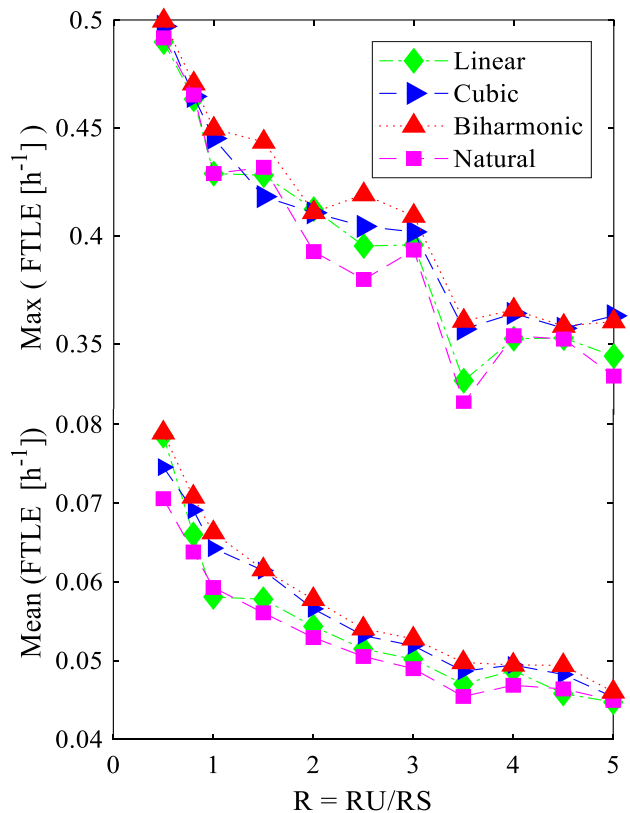

**Fig. 14:** Maximum and Mean FTLE calculation with different interpolation schemes for different grid resolution in Moreton Bay.

The PDF of the four different interpolation techniques in Moreton Bay for the full domain is shown in Fig. 15a. The result for the analytical kinematic models (Section 3.1.1) showed that the FTLE obtained from the velocity field with the biharmonic interpolation gives the closet value to the analytical model FTLE. As there is no direct structured velocity field data in Moreton Bay, here the biharmonic interpolation technique is considered as the structured data. The RMSE of the FTLE result of the three other interpolation schemes will use the FTLE obtained from the biharmonic scheme in Fig. 15(b). The results from Fig. 15(a) show that the ridges of the FTLE field marked are similar for the interpolation schemes. However, the peaks of the PDF for the natural and linear are significantly different from those of the biharmonic and cubic interpolation schemes. The localized effect, however, does not significantly affect the spatial average of the FTLE field.

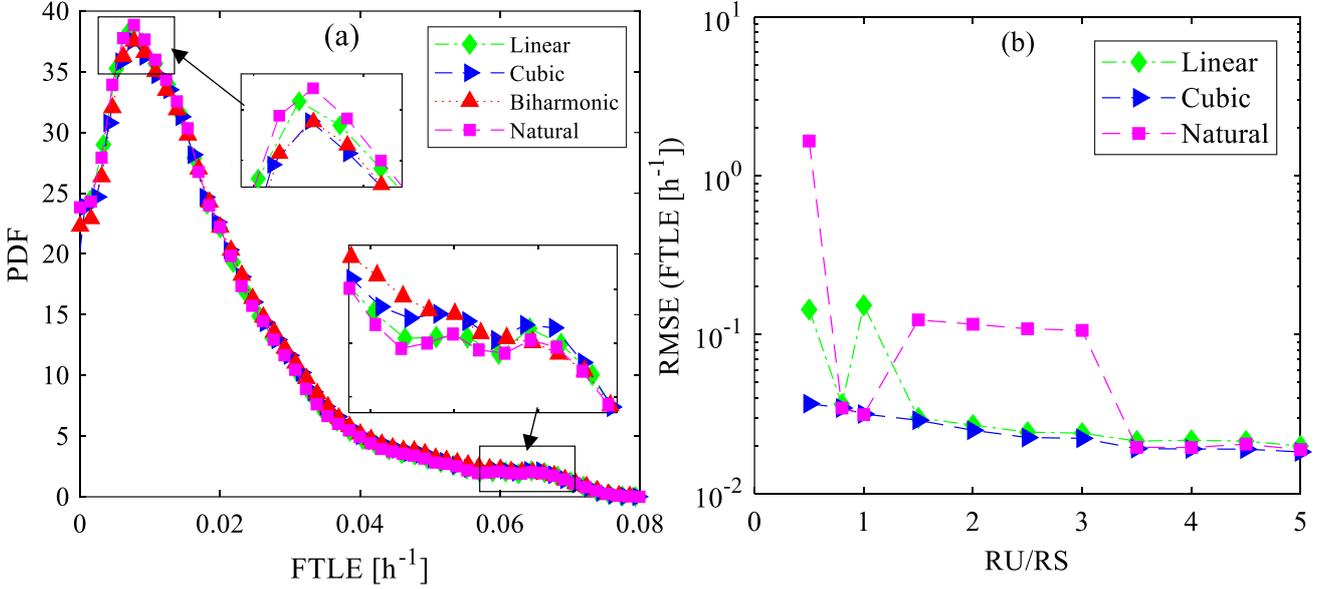

Fig. 15: Forward FTLE obtained in Moreton Bay for different interpolation schemes (a) PDF and (b) RMSE.

Fig. 15(b) shows the RMSE of Moreton Bay for the three interpolation schemes i.e., cubic, natural and linear. Here to examine the RMSE of Moreton Bay, the biharmonic scheme is considered as a structure grid. RMSE is calculated between the FTLE calculated using the biharmonic and another three interpolation schemes (linear, cubic and natural) which is plotted in Fig. 15(b). The levels of uncertainty in the estimated FTLE for natural and linear interpolation schemes are higher than that of the cubic interpolation scheme. This is expected as the linear and natural interpolation schemes have 1st order appromation while the cubic schemes have a higher order approximation. The comparison of the uncertainty in FTLE for the different interpolation schemes from Fig. 15(b), shows that the cubic scheme resulted in the minimum error for the range of resolution considered. Therefore, from the results it can be concluded that resolution selection is a far more important than the consideration of interpolation scheme selection from unstructured to structured grids. This is important for the flow and how the flow is controlled by the bathymetry area. So, for the process of conversion in field data, care should be taken to choose resolution and interpolation schemes to minimize the error. Otherwise, the wrong estimation of LCSs ridges and locations may arise in the result. This result is important for extracting LCS structures in coastal water system where complex bathymetry and boundaries imposed a constraint on modelling practices.

## 3.2     Effect of noise on LCSs

Hyperbolic and elliptic LCSs were calculated in this study to investigate the effect of random noise on their diagnostics. Noise was added to the true velocity fields for the kinematic models, doubles gyre and meandering jets as described in Section 2.3.3 and Equation (9). Both hyperbolic forward FTLE and elliptic LCSs were calculated for these analytical models. For the hyperbolic FTLE, the effect of noise on the ridge of maximum FTLE values, a proxy for the hyperbolic LCS, and the spatial averaged FTLE, a metric to characterise the mixing strength, is investigated. Similarly, the effect of the noise on the identification of closed orbits characterised by a constant tangential deformation gradient is investigated for the elliptic LCS. Because there are no long-lived closed orbits in Moreton Bay within the timescale captured by the model, only the effect of the noise on the hyperbolic FTLE field was investigated [29]. The structured velocity field obtained using biharmonic interpolation on the unstructured model output was used as the true model in Equation (9) for the Moreton Bay. This is because biharmonic interpolation produced the output closest to the true analytical models as discussed in Section 3.1.1. Because the standard deviations of the three flow fields are different, it should be noted that the resulting noise magnitudes are different. Specifically, the noise weighted by $k = 0 - 2$ correspond to a range of $0 – 65$ %, $0 – 16$ % and $0 – 185$ % for the double gyre, meandering jet and Moreton Bay cases, respectively.

The introduction of noise leads to an unsystematic change in the continuity of ridges of the maximum FTLE values. For example, Fig. 16 shows the FTLE fields for the double gyre and meandering jet for $k = 0$ and $k = 0.5$. While overall, the signature of the ridges is still discernible from visual inspection up to $k = 0.5$ - corresponding to 16.31 % & 4.19 % for the double gyre and meandering jet, respectively- the discontinuity has an implication on extracting continuous LCS lines from the ridges of FTLE in the presence of large random noise. This is due to the perturbation of the underlying flow field as a result of the superimposed noise. Ensembled average of more realizations of the FTLEs generated from degraded velocity could however reconstruct the FTLE ridges.

Furthermore, it should be noted that if the flow field from which the flow map is obtained is divergence free, the area preservation in the definition of the FTLE according to Equation (8) should be non-negative [58]. However, in addition to the inaccuracies resulting from the flow map integration, the noise in the flow field (Equation 9) introduces divergence to the system in a similar way that a large diffusion component can significantly distort the advective strength in an advection-diffusion system. This induced divergence resulted in nonzero FTLE values. A measure of the divergence, i.e., the area of the PDF for nonnegative FTLE field increases with the increase in the noise magnitude. Our preliminary analysis (not shown) of the level of divergence from the PDF of the FTLE at a given percentage of noise is the same for both the kinematic models and the Moreton Bay model output. This result suggests that the nonnegative FTLE field can be used as a measure of inaccuracies of the velocity field, consistent with the work of Beron-Vera, Olascoaga [58] in the West Florida Shelf.

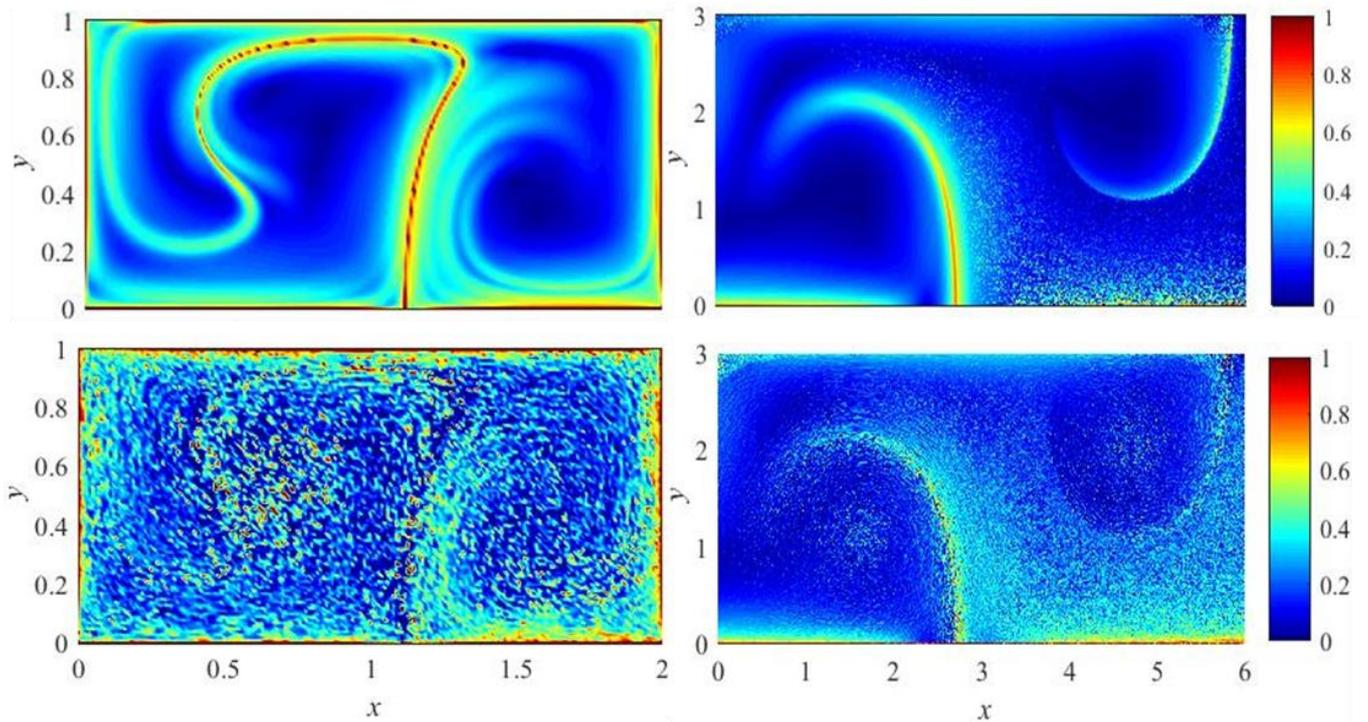

**Fig. 16:** Forward FTLE field for double gyre (left) and meandering jet (right); (top) True velocity field, i.e. $k = 0$; and (bottom) Degraded velocity field $k = 0.5$.

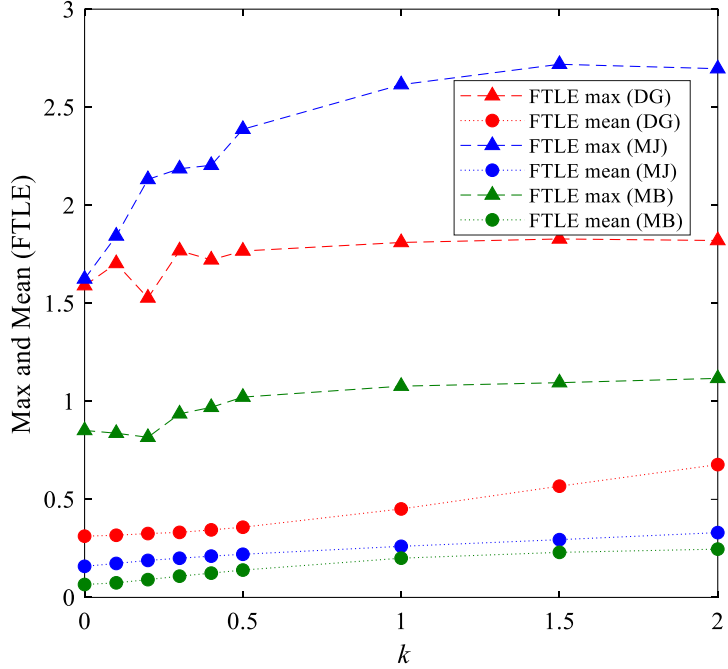

**Fig. 17:** Noise effect on the mean and maximum of the positive FTLE field for the double gyre (DG) and the meandering jet (MJ) and Moreton Bay (MB).

We investigated the spatial distribution of the absolute FTLE error, i.e., the difference between the FTLE obtained using the analytical flow field and the degraded velocity (not shown). It was found that the FTLE absolute error peaks near the ridges of the maximum FTLE values. These locations are, characterised with large dispersion or clustering which are the properties being identified by the FTLE and thus, associated with large velocity gradients and large velocity errors.

The noise effect on the mean and maximum of the positive FTLE field are presented in Fig. 17. The result showed that there are quantitative differences in the maximum and mean values of the FTLE for different values of $k$. For example, a slight increase in the noise level (e.g., $k = 0.1$, i.e., corresponding to 7 % in the double gyre and 11 % in the meandering jet maximum FTLE values), resulted in 3 % and 0.8 % of increase in the FTLE values for double gyre and meandering jet respectively. Beyond this $k$ value, the resulting maximum FTLE values are not significantly different for both analytical models (double gyre and meandering jet cases). There is no significant effect of the noise on the maximum values of the FTLE field for the Moreton Bay case even for 185 % added noise. Similarly, for all the three cases mean values of FTLE are not significantly affected for $k$ values up to 1 corresponding to 32 % noise for double gyre, 8 % noise for meandering jet and 86 % noise for Moreton Bay. The rapid decrease in the mean

FTLE values for $k > 1$ are associated with a large number of particles advected past the boundary with the increase in the noise level.

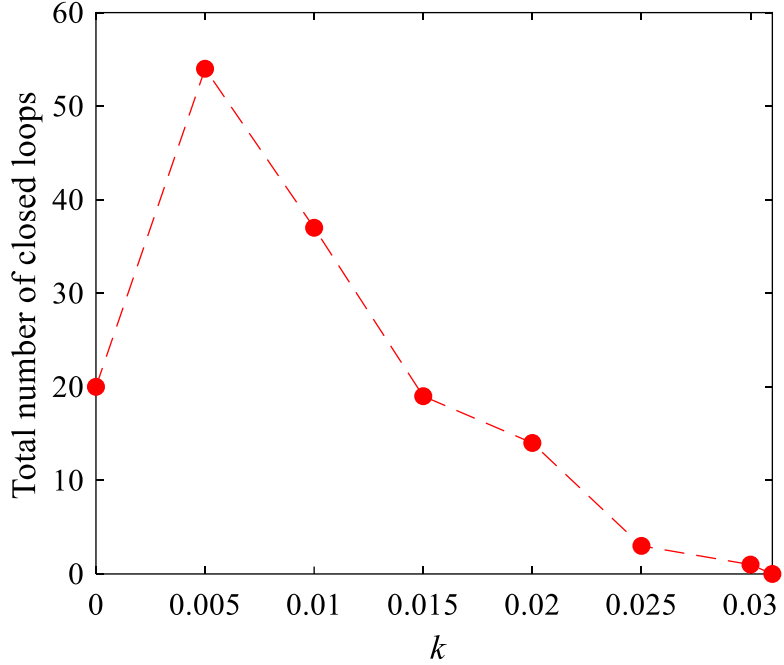

**Fig. 18:** Noise effect on elliptic LCSs of the double gyre kinematic model.

To investigate the effect of noise magnitude on the closed orbits of elliptic LCS, the double gyre kinematic model data is analysed and presented in Fig. 18. No closed orbits were found in the meandering jet and Moreton Bay. Closed orbits identified by a fixed stretching ratio, $\lambda$, have constant tangential deformation [48]. The effects of noise on the number of elliptic closed loops identified for a fixed stretching ratio, $\lambda$ ranging from 0.8 - 1.2 is examined and shown in Fig. 18. The range of $\lambda = 0.8 - 1.2$ with a fixed step of 20 was selected because the combination best captured the closed loops of the vortical gyre structures in the double gyre flow field. Fig. 18 shows that when the weighting factor increases to more than $k = 0.031$ corresponding to about 1 %, no closed loops could be further identified. This is consistent with the visual inspection of the ridges of maximum FTLE where addition of the random noise resulted in the discontinuity of the hyperbolic LCS due to the divergence in the underlying flow field similar to a diffusive system. On the other hand, there was no notable effect of random noise on the length of the identified outermost closed loops for the selected stretching ratio $\lambda$.

In summary, the noise effect on the maximum and mean FTLE values was not significant for different magnitudes of $k$. The implication is that the approximate locations of high values of FTLE as well as mixing strength are not significantly affected by the random noise. On the other hand, the identification of continuous hyperbolic LCS using the ridges of the FTLE field and closed orbit of elliptic LCS acting as a transport barrier can be significantly impaired by the presence of random noise analogous to a diffusive system due the divergence in the underlying flow field. Therefore, continuous hyperbolic LCS and closed orbits of elliptic LCS may be reconstructed from the flow map by considering formulations that take the random perturbation into account [59].

## 4. Conclusion

We have investigated the effects of mesh resolution, interpolation scheme and random noise distribution on LCSs for better understanding of data conversion from unstructured to structured grid data using two kinematic models and outputs of a hydrodynamic model. The results showed that the resolution of the velocity field grid is more important than the interpolation scheme for converting data from an unstructured to a structured grid. The errors resulting from grid resolution affects both the location and the magnitude of the FTLE fields. While the biharmonic and cubic interpolation schemes showed results closest to those from the true flow field, it was found that LCSs are not significantly affected by the level of inaccuracies resulting from interpolation schemes. In order to investigate the level of robustness of LCS to higher magnitude errors, the underlying flow fields were degraded with normally distributed random errors. Attributed to the divergence in the underlying flow field, the results showed that random errors in the order of 1-10 % break down the continuity in ridges of maximum FTLE and closed orbit elliptic LCS. It can be concluded that while peak values of FTLE and mixing strength measured by the FTLE are not affected by the noise, extraction of continuous transport barriers are highly impaired. The result has significant implications on the suitability of applying LCS formulations based on a deterministic flow field to diffusive coastal waters.

## Acknowledgements

We thank Professor H. Zhang and Dr. Y. Yu for access to hydrodynamic model and field data for Moreton Bay. The project is supported through Australia Research Council Linkage Project grant LP150101172 and Discovery Project grant DP190103379.